# Physics and science: the art of taking a stance about undecidable questions

Fabien Paillusson and Matthew Booth

School of Mathematics and Physics, University of Lincoln UK

20[th] Century physics has been struck by a series of shocking results that have raised issues with regards to what knowledge can be accessed about nature, what knowledge can be derived from our physical theories and, in general, has led to various debates about what physics actually is. It is tempting to think those existential crises in physics are fairly new. In this essay, we will argue that they have always existed, even before what we now call the scientific revolution. Instead of being a problem, proposing answers to apparently undecidable questions and asserting one's stance in spite of uncertainty could well be a defining feature of physics and science at large.

We shall illustrate and develop our claim by first reminding the history of one particular long-standing question in physics which has had, and still does have, a dramatic impact on our world view. We will then turn to more general notions of undecidability within physics and ask if there is any connection to similar concepts in computing and mathematics. Next, we will discuss the relationship between philosophical and scientific questions in the context of undecidability, before returning to some contemporary examples. Finally, we shall discuss the problem of consciousness in light of the thesis we develop herein.

## The Celestial Waltz: Who is Leading?

Most children are taught that, although the Sun appears to orbit the Earth, in actual fact it is the Earth that is orbiting the Sun. Inquisitive pupils may wonder how and why this inaccurate world view managed to last for more than two millennia! While it is easy to dismiss such questions by claiming that people then were just ignorant until the scientific revolution, such an answer does not do justice to the intellectual challenge that the pioneers of the scientific revolution had to overcome.

Let us convince ourselves with a simple model. In Fig. 1(a) we can clearly recognise the distinctive Keplerian paths drawn by the planets as they orbit the Sun. Likewise, the astronomers among us will recognise the equally distinctive retrograde motions of Mars about the Earth in Fig.1(b)[1]. When deciding between the heliocentric or the geocentric models' worldviews, it is tempting to think that we only need to point a telescope towards the sky in order to see the answer. However, Fig.1(c) and Fig.1(d) show that providing an answer to "which world view is real?" is not so simple. By removing the "trails" drawn by the planets along their trajectories, we see that the actual relative position (and in fact the relative motions too) are entirely equivalent in the two models. This is not surprising since the mathematical operation we have applied between the two Keplerian views is global and therefore affects the position of all bodies in the same manner, thus leaving invariant their relative motion.

---

[1] The trajectories of the geocentric model were obtained by applying a global change of coordinates to the heliocentric trajectories so as to pin the position of the Earth. In practice, the way it is done is by applying the parallelogram rule. Indeed, if $E$ stands for the Earth, $S$ for the Sun and $P$ for any other planet then $\overrightarrow{EP}(t) = \overrightarrow{ES}(t) + \overrightarrow{SP}(t) = \overrightarrow{SP}(t) - \overrightarrow{SE}(t)$ where the vectors $\overrightarrow{SP}$ are obtained from the heliocentric model.

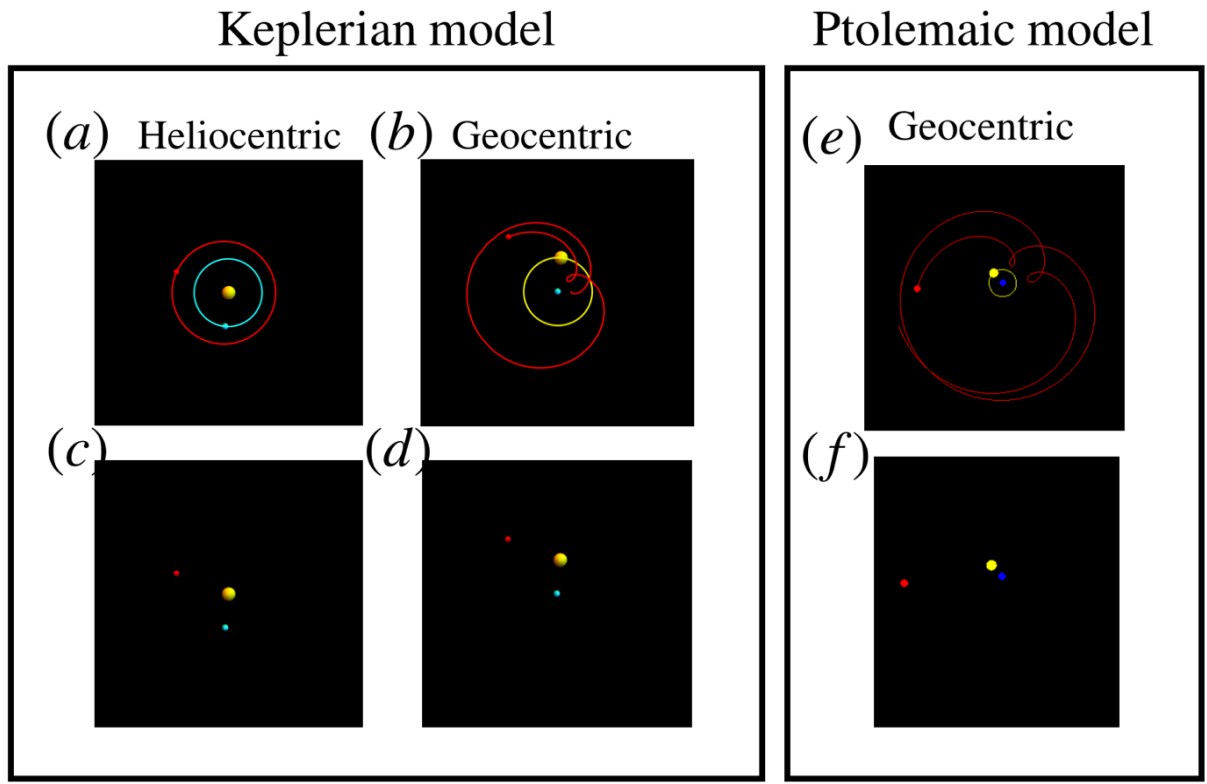

**Figure 1**: Computer generated heliocentric vs geocentric viewpoints of the motion of the Sun (Yellow), Earth (Blue) and Mars (Red). Trajectories in the geocentric model (b) are obtained by applying a coordinate transformation to the trajectories of the heliocentric model (a). Frames (c) and (d) are copies of frames (a) and (b) where the planets' path has been removed. Images (e) and (f) were obtained from Ref. [1] and represent a geocentric view as depicted in the Ptolemaic model using epicycles.

The fact that simply looking at the sky won't do, no matter how precise our measurements are and no matter how many celestial bodies we are able to observe, is called kinematic equivalence. The impossibility of deciding between two competing world views or theories, solely by measuring the quantities necessary for their description, illustrates what is called *underdetermination of theory by evidence* in the philosophy of science [2]. When two theories $T_1$ and $T_2$ are both consistent with the empirical evidence $E$, we are unable to decide between them:

$$\frac{\begin{array}{c} T_1 \text{ entails } E \\ T_2 \text{ entails } E \\ E \text{ is the case} \end{array}}{\text{Either } T_1 \text{ or } T_2 \text{ could be true or neither}}$$

In short, based on the description and measurement of motion alone, it is *empirically undecidable* to know whether the Sun is orbiting the Earth or vice versa. We can only say that either $T_1$ or $T_2$ could be true.

How can one then decide which theory is correct, or if either is correct at all? The dominant world view must be selected via the application of additional rules that supposedly grant the ability to decipher reality from illusion. In antiquity the emphasis was on sense-data and aesthetics. For example, from this perspective, the belief that the Earth is moving cannot be justified if we do not feel its motion. Consequently, Aristotle developed a cosmological model grounded in sense data, final causes and aesthetics. This was a geocentric world view with concentric spheres each containing one of the known planets.

Aristotle's model was eventually replaced by the much more successful geocentric model of Ptolemy, which kept much of the Aristotelean philosophy but supplemented it with new concepts; the idea of *epicycles* is central to his theory. The qualitative similarity between the Keplerian trajectories depicted in Fig.1(b) and the Ptolemaic trajectories depicted in Fig.1(e) is striking.

Fourteen centuries later Copernicus realised that a simpler model, which retained the aesthetic ideal of circular trajectories in the Heavens, was possible if the geocentric worldview was abandoned. Thus, despite having no empirical evidence (i.e. evidence grounded in sense-data) that Earth is moving about the Sun, Copernicus proposed arguments for why this was indeed the case. He argued that our senses might be liable and contended that simplicity must prevail above all other rules. Influential scholars such as Kepler and Galileo went on to contribute by further advocating for the rationality of the heliocentric view.

Then came Isaac Newton and his Principia. In his treaty on the mathematical principles of natural philosophy Newton does much more than simply stating his famous three laws of motion. On their own, these laws have a big flaw: while they rely on an absolute conception of space and time, it is *prima facie* undecidable to know which frames of reference are privileged enough to be in simple uniform rectilinear motion with respect to absolute space. The theoretical framework built by Newton in the first two books of Principia is like an arch in construction; the keystone, which gives the whole structure its integrity, is found in Book III where Newton discusses the Solar System. Here, Newton requires four additional rules of reasoning (*simplicity*, *universality of effects*, *universality of qualities* and *rational induction*). Combining these rules and his three laws of motion, he concludes that a) the Sun is almost identifiable as the centre of absolute space and that b) the planets, including the Earth, draw elliptic orbits which have their common focus at the centre of the Sun. This newly developed theoretical framework put the final nail in the coffin of geocentricism and has reigned supreme ever since.

**Fifty shades of undecidability in science**

The detailed example discussed in the previous section has shown that scholars were confronted with a form of undecidability, called underdetermination, with regards to the place of the Earth in the Universe. In the daily practice of science there is a further distinction to be made, however, between two forms of underdetermination: holistic [2] and contrastive [3].

Holistic underdetermination refers to underdetermination in response to failed predictions; one is unable to effectively isolate individual hypotheses from the 'web' of supporting assumptions and identify flaw(s) in an incorrect theory. A good example of holistic underdetermination is the failure of Newton's law of gravitation to accurately predict the orbit of Uranus during the 1$^{st}$ half of the 19$^{th}$ Century. Rather than rejecting Newton's model, Bouvard argued that there might instead be an unknown planet that perturbs Uranus. If hypothesis 1 ($H_1$) is Newton's law of gravitation, hypothesis 2 ($H_2$) is that there are 7 planets in the solar system, and $P$ refers to a prediction regarding the orbit of Uranus, the logical structure of this holistic underdetermination is:

$$\frac{(H_1 \text{ and } H_2) \text{ entails } P}{\text{Either } H_1 \text{ is false}, H_2 \text{ is false or both are false}}$$
$$\text{It is not the case that } P$$

In response to a failed prediction, there is no logical basis on which to differentiate the status of $H_1$ or $H_2$. In other words, not observing $P$ does not permit one to single out $H_1$ or $H_2$ as a false premise. The decision to discard one or both of the hypotheses must therefore be based on ampliative principles.

Contrastive underdetermination is a form of radical scepticism that calls into question any theory, no matter how consistent it is with empirical evidence, on the basis that there is a possibly infinite number of alternative theories which, "for want of anything analogous in our experience, our minds are unfitted to conceive". The geocentric view, for example, held its ground for about twenty centuries. While it would have been easy to infer that for a view such as this to last for so long it must be "true" in some absolute sense, it was eventually superseded by a heliocentric view justified by the Newtonian framework.

Looking more closely at scientific practice across the ages, different forms of "undecidability" pertaining to the validity of scientific claims have been pointed out by past thinkers. Identifying these different forms of undecidability can help us further appreciate which aspects of scientific practice are concerned with a form of undecidability.

Hume's problem of induction, for example, contends that the traditional view - that scientific laws are simply inferred from repeatable observations - cannot be logically valid. For this to be valid it is required that infinitely many repetitions of the same experiment are performed, which is impossible in a finite amount of time. A connection can then be proposed between contrastive underdetermination on the one hand and Hume's problem of induction on the other. Both are inextricably related to empirical observation and the logical invalidity of a form of inductive reasoning: in a theoretical sense for the former and in an empirical sense for the latter. Indeed, Hume's induction problem offers a new perspective about contrastive underdetermination. In order to answer a question such as 'is theory $T$ the best theory?', it is necessary to generalise about an entire class, possibly infinite, of objects (all theories except for $T$) based only on a limited number of comparisons. When Reichenbach said that the principle of induction was "the means whereby science decides upon truth", it seems he may have been correct in more ways than one. The issue here is that even if every theory tested so far against theory $T$ is deemed inferior to it, it is always possible that the next theory will be superior.

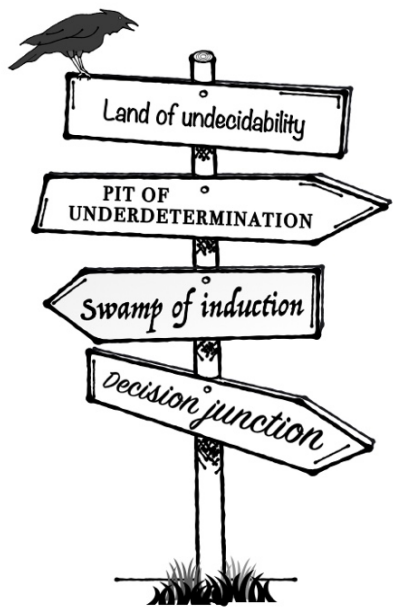

**Figure 2**: "If you don't know where you're going, any road can take you there."

Closer to our contemporary scientific practice, problems arise when making computer-based predictions for very large systems [4]. When bridging microscopic and macroscopic models of physical systems via the thermodynamic limit, a form of invalid induction occurs whereby one extrapolates a finite size or finite time behaviour to infinity. In absence of specific redundancies in the model, such strategies are a priori unjustified, and the model should be run on time and length scales so as to be as close as possible to the real system they seek to characterise. But even then, the halting problem states that there is no universal algorithm to determine whether such a computation is ever going to terminate on any given input (e.g. initial conditions); which can demonstrably lead to practical problems in theoretical physics [5]. Finally, algorithmic undecidability is closely related to the undecidability of questions which can be expressed within first order predicate logic: although this formal system is consistent and all true propositions in it are provable, checking with some automated procedure whether a given proposition is actually true is subject, in principle, to the same algorithmic undecidability as that described above. For example, some recent work [6] has shown that the band-gap problem of condensed-matter physics was both computationally and logically undecidable. The similarity of the questions "will the next step in an algorithm be terminal?" and "will the next theory be better than theory $T$?" suggests that the problem of induction is a point of contact between different notions of undecidability in the scientific practice. A question then remains: surrounded by all these forms of undecidability, how do scientists decide anything at all?

**The art of taking a stance**

In his book *The British Empiricists,* S. Priest defines a *philosophical question* as "one for which we have no method for answering". On this view, the task of philosophers is precisely to grapple with such undecidable questions and to suggest potential avenues for resolution. In the present essay we define a *scientific question* as a question deriving from a philosophical question once a method of resolution, usually involving mathematics, empirical knowledge and ampliative assumptions, has been proposed. In substance, what these ampliative principles do is transform an inductive inference into a deductive one, thus enabling closure. Scientists, in turn, are those tasked with answering such decidable scientific questions. Our main thesis in the essay is therefore that proposing an answer to undecidable questions becomes a defining feature of scientific practice, and that this is where philosophy and science interface. We believe this view opens-up new perspectives about the relation between science and philosophy, as we shall discuss below.

Based on the view we put forward, the famous statement by B. Russell that - *questions which are already capable of definite answers are placed in the sciences, while those only to which, at present, no definite answer can be given, remain to form the residue which is called philosophy* - was tautological; scientists give answers to scientific questions that are by definition outside of philosophy. What this means is that philosophical questions do play a role in the emergence of new scientific questions but more as a 'parent question' giving rise to 'daughter questions' than as a 'travelling question' moving from philosophy to science. Thus, a philosophical question always cohabits with its derived scientific questions, however prominent these scientific questions (and their answers) may be.

We have seen how Aristotle, Copernicus and Newton each proposed a different scientific question in response to the same philosophical question about the place of the Earth in the Universe. They were also acting as scientists in that they each provided answers to their own scientific recasting of the philosophical question, and each of them succeeded in convincing parts of their contemporaries that the proposed methods of resolution were better than the proposed alternatives. We claim that this process is an essential feature of science.

As another example we might consider the long-standing philosophical question of whether matter is discrete or continuous. Democritus, Parmenides, Lucretius and much later Dalton, Boltzmann and Einstein held an atomistic view of the world and developed convincing arguments accordingly. In parallel, Aristotle, Plato, and much later, Loschmidt, Mach and Ostwald developed views of matter based on continuous space-filling substances. Each of these thinkers had strong enough arguments to propose scientific questions, the answers to which would provide temporary closure on whether matter is particulate or not. It is also insightful to look at contemporary discussions on the subject. While we teach and celebrate atomism in the light of Einstein's 1905 seminal paper on Brownian motion, the world view suggested by Quantum Field Theory appeals to a reality of space-filing fields of which particles are but mere excitations [7]. This is a striking example of a philosophical question that has survived in spite of the many scientific paradigms it has engendered. Scientists perpetually reformulate a given philosophical question into scientific questions by taking a stance, and in doing so often provide closure, if only for a few generations.

The title of this essay implies there is an 'art' involved in taking a stance. This suggestion is intended to pay tribute to the inventiveness that is required to recast a philosophical question as a scientific one. The act of producing transformative art is always risky and the same might be said for science: the audacity required to take a stance, i.e. to make assumptions for which there are no empirical justifications, and to follow wherever these may lead, is necessary for scientific development.

We end this section by commenting on the status of mathematics relative to the sciences. While mathematics does contribute to many sciences, it is often debated whether or not it is a science in and of itself. On the view we have developed here, the very fact that conjectures are commonplace in mathematics, even though they may never be formally proven or disproven, is evidence that mathematicians are also confronted with undecidable questions and consequently do need to take a stance in their practice. In that sense, therefore, mathematics as a discipline does indulge in the same 'dark arts' as the sciences.

## Crisis or the new normal?

So far, we have developed the thesis that being confronted with undecidable philosophical questions and proposing decidable 'daughter questions', or scientific questions, is a defining feature of scientific practice. According to Kuhn, it is during the pre-paradigm or crisis stages of the development of a scientific discipline in which scientists tend to conceive of alternative theories: *"… since no experiment can be conceived without some sort of theory… It is, I think, particularly in periods of acknowledged crisis that scientists have turned to philosophical analysis as a device for unlocking the riddles of their field."*

In this view, each time a scientific discipline appears to be in crisis, it naturally reaches a 'new normal', where the principles on which its answers are based are reinvented from a philosophical perspective. This seems to fit perfectly with our description of the interplay between philosophy and science, whereby scientific questions are reformulated by the refocussing of philosophical questions during times of scientific crisis.

We shall quickly have a look at some of the 20th Century crises in physics, namely chaos and quantum mechanics and the kind of 'new normal' they have led to.

Although Newton's theory of gravitation predicted the orbits of individual planets such as Earth around the sun, as soon as a third body such as the moon was added, it became seemingly impossible to solve the equations explicitly within the Newtonian framework. Poincaré suggested that the three-body problem would require new mathematical tools that had yet to be invented, and Hadamard even suggested that such systems obeyed no laws at all. Much later, E. Lorenz found that sensitivity to initial conditions also occurred during the numerical simulations of simple dissipative systems. In principle, the initial state of a system is a point in some multidimensional 'phase' space, which in conjunction with the laws of mechanics entirely determines future states of that system. However, any empirical error in the estimation of the initial state prevents one from making reliable long-term predictions about the future state of the system. An even more problematic feature of such systems is that even if the initial state is known exactly, if it contains an irrational number (e.g. $\sqrt{2}$), then it is undecidable to know which machine precision will provide a trajectory that is the closest to the physical system that the model is trying to reproduce. The study and analysis of such systems, called chaotic systems, has opened-up new conceptual and mathematical perspectives in physics. Despite the fact that it is impossible to provide a reliable long-term prediction about one particular system, some regularities may emerge from the chaos: different trajectories of the same chaotic system may have the same fractal dimension or they may yield a sequence of pseudo-random numbers following a reproducible probability distribution. These developments have, for example, permitted the efficient computational study of thermodynamic equilibrium properties of complex systems, not by directly integrating their dynamical equations, but by using fictitious dynamical equations (e.g. Markov Chains) which sample the same probability distribution as that of the actual chaotic dynamics. In a sense, chaos is now seen as a regime that physical systems can be in, which carries its own set of laws to be discovered.

The famous debate between Einstein and Bohr regarding quantum mechanics presents itself as a good example where an apparently undecidable question on the nature of uncertainty in quantum mechanics (i.e. is it inherent or due to incomplete knowledge?) becomes decidable by adding ampliative principles. Einstein, Podolsky and Rosen (EPR), in their seminal 1935 paper, proposed a reframing of this question; by relying on a specific definition of local realism they reached a form of closure and concluded that quantum mechanics is incomplete. This paper was instrumental in driving Bell's argument that local "beables" only exist if certain inequalities are satisfied and the ensuing experimental works (*e.g.* Aspect's 1980s experiments). Adopting the following notation: *QM* = "Quantum Mechanics", *EPR* = "*QM* does not describe all elements of local reality" and *BI* = "Bell's inequalities hold", the chain of arguments reads:

$$\frac{\begin{array}{c}EPR \text{ entails } (BI \text{ and } (QM \text{ entails not } BI))\\ \text{Not } (BI \text{ and } (QM \text{ entails not } BI)) \text{ entails not } EPR \text{ [contrapositive]}\\ \text{Not } (BI \text{ and } (QM \text{ entails not } BI)) \text{ [Aspect's experiments]}\end{array}}{\text{Not } EPR}$$

This leads to the current view that quantum mechanics provides a complete description of reality and that the uncertainty of Quantum Mechanics is then inherent to "nature" so to speak.

We have argued that the process of rephrasing undecidable philosophical questions as decidable questions and proceeding to take a stance was a defining feature of science. In the next section we are going to see how this view may help illuminate a particularly contentious problem in the philosophy of mind.

## Is the ghost in the shell?

We finish this essay by discussing how the thesis we have developed in the previous sections can help us to appreciate the difficulties occurring in the debate about phenomenal consciousness. The so-called Hard Problem of consciousness put forward by D. Chalmers, among others, contends that providing answers to questions that are related to the nature of consciousness is beyond the scope of current scientific practice, particularly if current theories are considered to be the final say on any given question relating to experience. The main argument is that since consciousness is not currently acknowledged as a fundamental property of matter, and there does not seem to be any way to combine the currently accepted objective properties of matter that could lead to conscious subjective experience, there must be something missing. It is interesting to analyse Chalmers' position as stressing the philosophical nature of the question "what is consciousness?" in the sense defined above, i.e. a question for which there was no method for answering at the time it was coined. Certainly, many avenues have been proposed, but no specific method for answering it has achieved consensus. What we mean by 'method' here is a consistent set of principles or hypotheses, possibly new ones, which would allow one to conclude the existence of consciousness (or lack thereof) in a given system. We now look briefly at positions such as illusionism, substance dualism and panpsychism, in an attempt to analyse which of these might enable a scientific formulation of the Hard Problem.

Illusionism [8], proposes that there is no Hard Problem and that subjective experience is no more real than, say, a rainbow would be. In sum it is an illusion. On the surface, this position certainly takes a stance on the nature of consciousness but the stance itself appears to beg the question according to many observers [9]; to be the victim of an illusion actually feels "like something", which requires phenomenal consciousness in the first place.

Substance dualism contends that there is coexistence between a physical body and a soul; the former being governed by the laws of physics and the latter granting us subjective experience. Accounting for agency may require an interaction between these two substances which makes it a viable avenue for closure. But the set of principles required to make any such avenue decidable is still under construction as far as the authors are aware. For example, assuming interaction between a soul and a body may lead to deviations from our laws of physics, for example the conservation of energy [10]. The issue, though, is that these deviations, if ever seen, must be identified as evidence for an interaction between a soul and a body rather than some "new physics" not involving any soul. We are then back to contrastive underdetermination. Therefore, ampliative principles are still needed in order to generate scientific questions in the sense defined in this essay.

Panpsychism contends that phenomenal consciousness is not an exclusive feature of human beings or even living organisms but that there could be some degree of consciousness in all physical systems a priori [11]. Some strong forms of panpsychism suggest that electrons have some degree of consciousness [12] while other versions grant consciousness based on some additional criteria. Panpsychism alone, however, is not enough to develop scientific 'daughter' questions; as is the case for substance dualism, further principles are required.

Functionalism is an example of such an additional principle in that it postulates the existence of some level of consciousness depending on the extent to which some objective function(s) can be emulated by a given system. Such a view appears to us as the closest to what we call a scientific question. Tononni's Integrated Information Theory for example posits that a system which has some ability to take information as an input, 'integrate' it and then respond according to a cause-effect or 'conceptual' structure, must be assumed to have some degree of conscious experience determined by the level of information integration being displayed. In a similar vein, Tegmark suggests that degrees of consciousness may be associated with the structural pattern of communication channels in a system. In light of the thesis developed in this essay, these positions constitute viable methods for science to answer some questions about the nature of consciousness, thereby providing temporary closure to the philosophical questions for which we would have substituted their scientific translation.

## Closing comments

In this essay, we have discussed the problems of uncertainty and undecidability faced by physics and most sciences. We have developed the idea that, far from being a new phenomenon, the existence of apparently unsolvable questions can be found in many places in the history of physics. We have also extended the discussion to show that most undecidability issues that the sciences are confronted with tend to revolve around a form of logically invalid induction. Finally, using supportive examples, we suggest that the creative frameworks proposed to address undecidable problems are in fact a defining feature of science, whereby what we called philosophical questions give birth to scientific questions that are decidable.

In closing, we propose a quote [13] which we believe captures the impossible task of science which we have defended in this essay:

> "*When Chuck Norris makes an inductive inference, it becomes deductive*".